\documentclass[a4paper,UKenglish]{lipics}

\theoremstyle{plain}
\newtheorem{proposition}[theorem]{Proposition}

\usepackage{hyperref}
\renewcommand\url[1]{\href{#1}{#1}}
\usepackage{microtype}
\usepackage{bussproofs}
\usepackage{stmaryrd}
\usepackage{xcolor}
\usepackage{MnSymbol}

\title{Termination of $\lambda\Pi$ modulo rewriting using the size-change principle (work in progress)}
\titlerunning{Termination of $\lambda\Pi$ modulo rewriting using the
  size-change principle}

\author[1,2]{Fr\'ed\'eric Blanqui}
\author[1,2,3]{Guillaume Genestier}
\affil[1]{LSV, ENS Paris-Saclay, CNRS, Université Paris-Saclay}
\affil[2]{Inria}
\affil[3]{MINES ParisTech, PSL University}
\authorrunning{F. Blanqui and G. Genestier}

\Copyright{Fr\'ed\'eric Blanqui and Guillaume Genestier}


\lstdefinelanguage{Dedukti}
{
  alsoletter={=->:\#},
  keywords={Type,def,-->,->,=>,:=,:,.,\#SNF,\#NAME,\#PRINT},
  delim=[s][\color{brown}]{\[}{\]},
  comment=[n]{(;}{;)},
  string=[b]{"},
  stringstyle=\color{orange},
  commentstyle=\color{red},
  showstringspaces=false
}
\definecolor{grispale}{RGB}{240,240,240}
\let\<\leqslant

\DeclareMathOperator\Type{Type}
\DeclareMathOperator\Kind{Kind}
\DeclareMathOperator\arity{ar}

\DeclareMathOperator\Nat{Nat}
\DeclareMathOperator\SCT{SCT}
\DeclareMathOperator{\CC}{CC}

\newcommand\call{~\widetilde\succ~}

\definecolor{vert}{RGB}{0,150,0}

\newcommand\hide[1]{}

\newcommand\moins\setminus
\newcommand\vide\emptyset

\renewcommand\a\longrightarrow
\newcommand\A\Rightarrow
\renewcommand\aa\longleftrightarrow
\renewcommand\AA\Leftrightarrow
\newcommand\la\leftarrow
\newcommand\lA\Leftarrow
\newcommand\ad\downarrow
\newcommand\Ad\Downarrow
\newcommand\au\uparrow
\newcommand\Au\Uparrow
\renewcommand\to\mapsto

\newcommand\ab{\a_\b}

\newcommand\ar{\a_\cR}

\newcommand\I[1]{\llbracket{#1}\rrbracket}

\newcommand\ex\exists
\newcommand\all\forall
\newcommand\ou\vee
\newcommand\bigou\bigvee
\newcommand\biget\bigwedge
\newcommand\et\wedge
\newcommand\non\neg
\newcommand\B\Box
\renewcommand\th\vdash

\newcommand\sle\subseteq
\newcommand\sge\supseteq
\newcommand\slt\subset
\newcommand\sgt\supset
\newcommand\tle\unlhd
\newcommand\tge\unrhd
\newcommand\tlt\lhd
\newcommand\tgt\rhd
\newcommand\cle\preceq
\newcommand\cge\succeq
\newcommand\clt\prec
\newcommand\cgt\succ
\newcommand\qle\sqsubseteq
\newcommand\qge\sqsupseteq
\newcommand\qlt\sqsubset
\newcommand\qgt\sqsupset

\newcommand\lex{\mr{lex}}

\renewcommand\prod{\mr{prod}}
\newcommand\stat{\mr{stat}}

\newcommand\al\alpha
\renewcommand\b\beta
\newcommand\g\gamma
\newcommand\G\Gamma
\renewcommand\d\delta
\newcommand\D\Delta
\newcommand\ep\epsilon
\newcommand\vep\varepsilon
\newcommand\z\zeta
\renewcommand\t\theta
\newcommand\T\Theta
\newcommand\io\iota
\newcommand\kap\kappa
\renewcommand\l\lambda
\renewcommand\L\Lambda
\renewcommand\r\rho
\newcommand\s\sigma
\renewcommand\S\Sigma
\newcommand\up\upsilon
\newcommand\Up\Upsilon
\newcommand\vphi\varphi
\newcommand\w\omega
\newcommand\W\Omega

\newcommand\bs\boldsymbol
\newcommand\mi\mathit
\newcommand\mc\mathcal
\newcommand\mt\mathtt
\newcommand\mr\mathrm
\newcommand\mb\mathbb 
\newcommand\mg\mathbf
\newcommand\mk\mathfrak
\newcommand\ms\mathsf
\DeclareMathAlphabet\mz{OT1}{pzc}{m}{it} 

\newcommand\cC{\mc{C}}

\newcommand\cF{\mc{F}}

\newcommand\cR{\mc{R}}

\newenvironment{rul}
  {$\begin{array}{rcl}}
  {\end{array}$}

\newenvironment{rewc}[1][~~\a~~]
  {\begin{center}\begin{rew}[#1]}
  {\end{rew}\end{center}}

\hide{

{\theorembodyfont{\rmfamily} 

}}

\newenvironment{lstgeneric}[2]
  {\begin{list}{#1}{\topsep=.5mm\itemsep=.5mm\parsep=0mm%
    \itemindent=-3ex\labelsep=1ex\labelwidth=0ex #2}}
  {\end{list}}

\begin{document}

\maketitle

\begin{abstract}
The Size-Change Termination principle was first introduced to study the
termination of first-order functional programs. In this work, we show
that it can also be used to study the termination of higher-order
rewriting in a system of dependent types extending LF.
\end{abstract}

\section{Introduction}

The Size-Change Termination principle (SCT) was first introduced by Lee, Jones and
Ben Amram \cite{lee01popl} to study the termination of first-order
functional programs. It proved to be very effective, and a few
extensions to typed $\lambda$-calculi and higher-order rewriting were
proposed.

In his PhD thesis \cite{wahlstedt07phd}, Wahlstedt proposes one for
proving the termination, in some presentation of Martin-Löf's type
theory, of an higher-order rewrite system $\cR$ together with the
$\beta$-reduction of $\lambda$-calculus. He proceeds in two
steps. First, he defines an order, the instantiated call relation, and
proves that ${\a}={\ar\cup\ab}$ terminates on well-typed terms
whenever this order is well-founded. Then, he uses SCT to eventually
show the latter.

However, Wahlstedt's work has some limitations. First, it only
considers weak normalization, that is, the mere existence of a normal
form. Second, it makes a strong distinction between ``constructor''
symbols, on which pattern matching is possible, and ``defined''
symbols, which are allowed to be defined by rewrite rules.
Hence, it cannot handle all the systems that one can define in the
$\lambda\Pi$-calculus modulo rewriting, the type system implemented in
Dedukti \cite{assaf16draft}.

Other works on higher-order rewriting do not have those restrictions,
like \cite{blanqui05mscs} in which strong normalization (absence of
infinite reductions) is proved in the calculus of constructions by
requiring each right-hand side of rule to belong to the Computability
Closure ($\CC$) of its corresponding left-hand side.

In this paper, we present a combination and extension of both approaches.

\section{The $\lambda\Pi$-calculus modulo rewriting}
\label{LambdaPi}

We consider the $\lambda\Pi$-calculus modulo rewriting
\cite{assaf16draft}. This is an extension of Automath\hide{
  \cite{debruijn68sad}}, Martin-Löf's type theory\hide{
  \cite{martinlof84book}} or LF\hide{ \cite{harper93jacm}}, where
functions and types can be defined by rewrite rules, and where types
are identified modulo those rules and the $\beta$-reduction of
$\lambda$-calculus.

Assuming a signature made of a set $\cC_T$ of type-level constants, a
set $\cF_T$ of type-level definable function symbols, and a set
$\cF_o$ of object-level function symbols, terms are inductively
defined into three categories as follows:

\begin{center}
  \begin{tabular}{rr@{~::=~}ll}
  \mbox{kind-level terms} & $K$ & $\Type\mid(x:U)\rightarrow K$\\
  \mbox{type-level terms} & $T,U$ & $\lambda x:U.\,T\mid(x:U)\rightarrow T\mid U\,t\mid D\mid F$& where $D\in\cC_T$ and $F\in\cF_T$\\
  \mbox{object-level terms} & $t,u$ & $x\mid\lambda x:U.\,t\mid t\,u\mid f$& where $f\in\cF_o$\\
  \end{tabular}
\end{center}

By $\bar t$, we denote a sequence of terms $t_1\dots t_n$ of length
$|\bar t|=n$.

Next, we assume given a function $\tau$ associating a kind to every
symbol of $\cC_T$ and $\cF_T$, and a type to every symbol of $\cF_o$. If
$\tau(f)=(x_1:T_1)\rightarrow\dots\rightarrow(x_n:T_n)\rightarrow U$
with $U$ not an arrow, then $f$ is said of arity $\arity(f)=n$.

An object-level function symbol $f$ of type
$(x_1:T_1)\rightarrow\dots\rightarrow(x_n:T_n)\rightarrow D\,u_1\dots
u_{\arity(D)}$ with $D\in\cC_T$ and every $T_i$ of the form
$E\,v_1\dots v_{\arity(E)}$ with $E\in\cC_T$ is called a {\em
  constructor}. Let $\cC_o$ be the set of constructors.

Terms built from variables and constructor application only are called
\emph{patterns}:\\
$p ::= x\mid c\,p_1\dots p_{\arity(c)} \text{ where }c\in\cC_o$.

Next, we assume given a set $\cR$ of rewrite rules of the form $f\,
p_1\dots p_{\arity(f)} \a r$, where $f$ is in $\cF_o$ or
$\cF_T$, the $p_i$'s are patterns and
$r$ is $\beta$-normal. Then, let ${\a}={\ab\cup\ar}$ where $\ar$ is
the smallest rewrite relation containing $\cR$.

Note that rewriting at type level is allowed. For instance, we can
define a function taking a natural number $n$ and returning
$\Nat\rightarrow\Nat\rightarrow\dots\rightarrow\Nat$ with as many
arrows as $n$. In Dedukti syntax, this gives:
\begin{lstlisting}
def F : Nat -> Type.
[]  F 0     --> Nat.
[n] F (S n) --> Nat -> (F n).
\end{lstlisting}

\hide{
However, in order to interpret type symbols, we add the restriction
that, if $r$ is of the form $(x:T)\rightarrow U$, then $T$ contains no
defined symbol.  It seems difficult to avoid such a restriction in
interpretation mapping every type with a term set like the one we are
studying here.  However, it would be great to find a way to relax this
constraint, since it is quite common, especially in Dedukti, to encode
a logic in the $\lambda\Pi$-calculus modulo theory and in this case,
the decoding of a product does not respect this constraint.
}

Well-typed terms are defined as in LF, except that
types are identified not only modulo $\b$-equivalence but modulo
$\cR$-equivalence also, by adding the following type conversion rule:

\begin{prooftree}
  \AxiomC{$\G\th t:A$}
  \AxiomC{$\G\th A:s$}
  \AxiomC{$\G\th B:s$}
  \RightLabel{$\quad\text{if }A\aa^* B\text{ and }s\in\{\Type,\Kind\}$}
  \TrinaryInfC{$\G\th t:B$}
\end{prooftree}

Convertibility of $A$ and $B$, $A\aa^* B$, is undecidable in
general. However, it is decidable if $\a$ is confluent and terminating. So, a
type-checker for the $\lambda\Pi$-calculus modulo $\aa^*$, like
Dedukti, needs a criterion to decide termination of $\a$. This is the
reason of this work.

To this end, we assume that $\a$ is confluent and preserves typing.

There exist tools to check confluence, even for higher-order rewrite
systems, like CSI\^{}ho or ACPH. The difficulty in presence of
type-level rewrite rules, is that we cannot assume termination to show
confluence since we need confluence to prove
termination. Still, there is a simple criterion in this case:
orthogonality \cite{oostrom94lfcs}.

Checking that $\a$ preserves typing is undecidable too (for $\ab$
alone already), and often relies on confluence except when type-level
rewrite rules are restricted in some way
\cite{blanqui05mscs}. Saillard designed and implemented an heuristic
in Dedukti \cite{saillard15phd}.

Finally, note that constructors can themselves be defined by rewrite
rules. This allows us to define, for instance, the type of integers
with two constructors for the predecessor and successor,
together with the rules stating that they are inverse of each
other. \hide{In Dedukti syntax, this gives:

\begin{lstlisting}
Int : Type.

0 : Int.
def S : Int -> Int.
def P : Int -> Int.

[x] S (P x) --> x.
[x] P (S x) --> x.
\end{lstlisting}}

\section{The Size-Change Termination principle}

Introduced for first-order functional programming languages by Lee,
Jones and Ben Amram \cite{lee01popl}, the SCT is a simple but
powerful criterion to check termination. We recall hereafter the
matrix-based presentation of SCT by Lepigre and Raffalli
\cite{lepigre17draft}.

\begin{definition}[Size-Change Termination principle]
  The (strict) {\em constructor subterm relation} $\tlt$ is the
  smallest transitive relation such that $t_i\tlt c\,t_1\dots t_n$
  when $c\in\cC_o$.
  
  We define the \emph{formal call relation} by $f\,\bar p$
  $>_{call}g\,\bar t$ if there is a rewrite rule
  $f\,\bar p\a r\in\mathcal{R}$ such that $g\in\cF_T\cup\cF_o$ and
  $g\,\bar t$ is a subterm of $r$ with $|\bar t|=\arity(g)$.
  
  From this relation, we construct a \emph{call graph} whose nodes are
  labeled with the defined symbols. For every call $f\,\bar
  p>_{call}g\,\bar t$, an edge labeled with the \emph{call matrix}
  $(a_{i,j})_{i\<\arity(f),j\<\arity(g)}$ links the nodes $f$ and $g$,
  where $a_{i,j}=-1$ if $t_j\tlt p_i$, $a_{i,j}=0$ if $t_j=p_i$, and
  $a_{i,j}=\infty$ otherwise.

  A set of rewrite rules $\cR$ satisfies the \emph{size-change
  termination principle} if the transitive closure of the call graph (using the
  max-plus semi-ring to multiply the matrices) is such that
  all arrows linking a node with itself are labeled with a matrix
  having at least one $-1$ on the diagonal.
\end{definition}

The formal call relation is also called the dependency pair relation
\cite{arts00tcs}.

\hide{Let illustrate this notion with the Peano-like definition of addition
and multiplication:

\begin{center}
\lstset{backgroundcolor = \color{white}, language={Dedukti}}
   \begin{tabular}{|l|c|}
    \hline
\hide{    \begin{lstlisting}
def plus : Nat -> Nat -> Nat.
    \end{lstlisting}&\\}
    \begin{lstlisting}
[n]   plus 0     n --> n.
    \end{lstlisting}& $\vide$\\\hline
    \begin{lstlisting}
[m,n] plus (S m) n --> S (plus m n).
    \end{lstlisting}&
                      $\left\{\begin{pmatrix}-1&\infty\\\infty&0\end{pmatrix}\right\}$\\
    \hline
\hide{    \begin{lstlisting}
def mult : Nat -> Nat -> Nat.
    \end{lstlisting}&\\}
    \begin{lstlisting}
[]    mult 0     _ --> 0
    \end{lstlisting}& $\vide$\\\hline
    \begin{lstlisting}
[m,n] mult (S m) n --> plus n (mult m n).
    \end{lstlisting}&
    $\left\{
    \begin{pmatrix}\infty&\infty\\0&\infty\end{pmatrix},
    \begin{pmatrix}-1&\infty\\\infty&0\end{pmatrix}
    \right\}$\\
    \hline
  \end{tabular}
\lstset{backgroundcolor = \color{grispale}, language={Dedukti}}
\end{center}

In the last rule, the function \texttt{mult} calls \texttt{mult}
itself but also \texttt{plus}, that is the reason why two matrices are
associated to this rule.}

\hide{We must notice here that the $\SCP$ is a global criterion in the sense
that two rules can satisfy the SCP alone, but not when they are put
together. This property sounds quite natural since termination is
well-known to be non-modular.}

\hide{It would be tempting to replace the constructor subterm order used in the
construction of the call matrices by some other well-founded order.
However, Thiemann and Giesl showed \cite{thiemann05aaecc} that
there exist well-founded orders for which SCP does not imply
termination.}

\section{Wahlstedt's extension of SCT to Martin-Löf's Type Theory}

The proof of weak normalization in Wahlstedt's thesis uses an
extension to rewriting of Girard's notion of reducibility candidate
\cite{girard88book}, called computability predicate here. This
technique requires to define an interpretation of every type $T$ as a
set of normalizing terms $\I{T}$ called the set of {\em computable}
terms of type $T$. Once this interpretation is defined, one shows that
every well-typed term $t:T$ is {\em computable}, that is, belongs to the
interpretation of its type: $t\in\I{T}$, ending the normalization
proof. To do so, Wahlstedt proceeds in two steps. First, he shows that
every well-typed term is computable whenever all symbols are
computable. Then, he introduces the following relation which, roughly
speaking, corresponds to the notion of minimal chain in the DP framework
\cite{arts00tcs}:

  
\begin{definition}[Instantiated call relation]
  Let $f\,\bar t\call g\,\bar v$ if there exist $\bar p$, $\bar u$ and
  a substitution $\gamma$ such that $\bar t$ is normalizing,
  $\bar t\a^*\bar p\gamma$, $f\,\bar p>_{call}g\,\bar u$ and
  $\bar u\gamma=\bar v$.
\end{definition}

\noindent
and proves that all symbols are computable if $\call$ is well-founded:

\begin{lemma}[{\cite[Lemma 3.6.6, p. 82]{wahlstedt07phd}}]
\label{WKeyLemma}
If $\call$ is well-founded\hide{ and all the types occurring in the signature
are computable}, then all symbols are computable.
\end{lemma}

Finally, to prove that $\call$ is well-founded, he uses SCT:

\begin{lemma}[{\cite[Theorem 4.2.1, p. 91]{wahlstedt07phd}}]
  $\call$ is well-founded whenever the set of rewrite rules satisfies
  $\SCT$.
\end{lemma}

Indeed, if $\call$ were not well-founded, there would be an infinite
sequence $f_1\,\bar t_1\call f_2\,\bar t_2\call\dots$, leading
to an infinite path in the call graph which would visit infinitely often
at least one node, say $f$. But the matrices labelling the looping edges
in the transitive closure all contain at least one $-1$ on
the diagonal, meaning that there is an argument of $f$ which strictly
decreases in the constructor subterm order at each cycle. This would
contradict the well-foundedness of the constructor subterm order.

However, Wahlstedt only considers weak normalization of orthogonal
systems, in which constructors are not definable. There exist
techniques which do not suffer those restrictions, like
the Computability Closure.

\section{Computability Closure}

The Computability Closure ($\CC$) is also based on an extension of Girard's
computability predicates \cite{blanqui16tcs}, but for strong
normalization. The gist of $\CC$ is, for every left-hand side of a rule
$f\,\bar l$, to inductively define a set $\CC_\qgt(f\,\bar l)$ of terms
that are computable whenever the $l_i$'s so are. Function applications
are handled through the following rule:
\begin{center}
    \AxiomC{$f\,\bar l\qgt g\,\bar u$}
    \AxiomC{$\bar u\in\CC_\qgt(f\,\bar l)$}
    \BinaryInfC{$g\,\bar u\in\CC_\qgt(f\,\bar l)$}
    \DisplayProof
\end{center}
where ${\qgt}={(\succ_\cF,(\tgt\cup\a)_{\stat})_\lex}$ is a
well-founded order on terms $f\,\bar t$ such that $\bar t$ are
computable, with $\succ_\cF$ a precedence on function symbols and
$\stat$ either the multiset or the lexicographic order extension,
depending on $f$.

Then, to get strong normalization, it suffices to check that, for
every rule $f\,\bar l\a r$, we have $r\in\CC_\qgt(f\,\bar l)$. This is
justified by Lemma 6.38 \cite[p.85]{blanqui05mscs} stating that all
symbols are computable whenever the rules satisfy $\CC$, which looks like
Lemma \ref{WKeyLemma}. It is proved by induction on
${\qgt}$. By definition, $f\,\bar t$ is computable if,
for every $u$ such that $f\,\bar t\a u$, $u$ is computable. There are
two cases. If $u=f\,\bar t'$ and $\bar t\a\bar t'$, then we
conclude by the induction hypothesis. Otherwise, $u=r\,\gamma$
where $r$ is the right-hand side of a rule whose left-hand side is of
the form $f\,\bar l$. This case is handled by induction on the proof
that $r\in\CC_\qgt(f\,\bar l)$.

So, except for the order, the structures of the proofs are very similar in both
works. This is an induction on the order, a case distinction and, in
the case of a recursive call, another induction on a refinement of the
typing relation, restricted to $\beta$-normal terms in Wahlstedt's
work and to the Computability Closure membership in the other one.

\section{Applying ideas of Computability Closure in Wahlstedt's criterion}

We have seen that each method has its own weaknesses: Wahlstedt's SCT
deals with weak normalization only and does not allow pattern-matching
on defined symbols, while $\CC$ enforces mutually defined functions to
perform a strict decrease in each call.

We can subsume both approaches by combining them and replacing in the
definition of $\CC$ the order $\qgt$ by the formal call relation:
\begin{center}
    \AxiomC{$f\,\bar l>_{call} g\,\bar u$}
    \AxiomC{$\bar u\in\CC_{>_{call}}(f\,\bar l)$}
    \BinaryInfC{$g\,\bar u\in\CC_{>_{call}}(f\,\bar l)$}
    \DisplayProof
\end{center}
We must note here that, even if $>_{call}$ is defined from the
constructor subterm order, this new definition of $\CC$ does not enforce
an argument to be strictly smaller at each recursive call, but only
smaller or equal, with the additional constraint that any looping
sequence of recursive calls contains a step with a strict decrease,
which is enforced by SCT.

\begin{proposition}
  Let $\cR$ be a rewrite system such that ${\a}={\ab\cup\ar}$ is
  confluent and preserves typing. If $\cR$ satisfies $\CC_{>_{call}}$ and $\SCT$,
  then $\a$ terminates on every term typable in the $\l\Pi$-calculus
  modulo $\aa^*$.
\end{proposition}

Note that $\CC_{>_{call}}$ essentially reduces to checking that the
right-hand sides of rules are well-typed which is a condition that is
generally satisfied.

The main difficulty is to define an interpretation for types and type symbols
that can be defined by rewrite rules. It requires to use
induction-recursion \cite{dybjer00jsl}. Note that the well-foundedness
of the call relation $\call$ is used not only to prove
reducibility of defined symbols, but also to ensure that the
interpretation of types is well-defined.

If we consider the example of integers mentioned earlier and define
the function erasing every constructor using an auxiliary function, we
get a system rejected both by Wahlstedt's criterion since $\mathtt{S}$
and $\mathtt{P}$ are defined, and by the $CC_\qgt$ criterion since there is
no strict decrease in the first rule. On the other hand, it is
accepted by our combined criterion.
\begin{lstlisting}
Int : Type.                      0 : Int.
def S : Int -> Int.              def P : Int -> Int.

[x] S (P x) --> x.               [x] P (S x) --> x.
[x] returnZero x --> aux x.      []  aux 0     --> 0.
[x] aux (S x) --> returnZero x.  [x] aux (P x) --> returnZero x.
\end{lstlisting}

\section{Conclusion}

We have shown that Wahlstedt's thesis \cite{wahlstedt07phd} and the
first author's work \cite{blanqui05mscs} have strong
similarities. Based on this observation, we developed a combination of
both techniques that strictly subsumes both approaches.

This criterion has been implemented in the type-checker Dedukti
\cite{assaf16draft} and gives promising results, even if automatically
proving termination of expressive logic encodings remains a
challenge. The code is available at
\url{https://github.com/Deducteam/Dedukti/tree/sizechange}.

Many opportunities exist to enrich our new criterion. For instance,
the use of an order leaner than the strict constructor subterm for
SCT, like the one defined by Coquand \cite{coquand92types} for
handling data types with constructors taking functions as
arguments. This question is studied in the first-order case by
Thiemann and Giesl \cite{thiemann05aaecc}.

Finally, it is important to note the modularity of Wahlstedt's
approach. Termination is obtained by proving 1) that all terms
terminate whenever the instantiated call relation is well-founded, and
2) that the instantiated call relation is indeed
well-founded. Wahlstedt and we use SCT to prove 2) but it should be
noted that other techniques could be used as well. This opens the
possibility of applying to type systems like the ones implemented in
Dedukti, Coq or Agda, techniques and tools developed for proving the
termination of DP problems.

\smallskip
{\bf Acknowledgments.} The authors thank Olivier Hermant for his
comments, as well as the anonymous referees.


\end{document}